\documentclass[12pt]{article}
\usepackage[table, dvipsnames]{xcolor}
\usepackage{subcaption}
\usepackage{setspace}
\usepackage{calc,graphics}
\usepackage{changepage}
\usepackage[utf8]{inputenc}
\usepackage{textcomp,marvosym}
\usepackage{fixltx2e}
\usepackage{amsmath,amssymb}
\usepackage{cite}
\usepackage{nameref,hyperref}
\usepackage[right]{lineno}
\usepackage{microtype,todonotes}
\usepackage{acronym}
\usepackage{times}
\usepackage{bm}
\usepackage{comment}
\usepackage[plain,noend]{algorithm2e}
\usepackage{amsgen,amsmath,amssymb,latexsym}
\usepackage{graphicx}
\usepackage{enumerate}
\usepackage{type1cm}
\usepackage{xspace}
\usepackage{tikz}
\usepackage{float,lscape}
\usepackage{longtable}
\usepackage{makeidx}
\usepackage{multirow}
\usepackage{multicol}
\usepackage{footnote}
\usepackage{rotating}
\usepackage{graphics}
\usepackage{placeins} 
\usepackage{microtype}
\usepackage{geometry}
\usepackage{multirow}
\usepackage{booktabs}
\usepackage{pdfpages}
\usepackage{times}
\usepackage{comment}
\usepackage[plain,noend]{algorithm2e}
\usepackage{arydshln}
\usepackage{amsmath}
\usepackage{type1cm}
\usepackage{tikz}
\usepackage{color}
\usepackage{rotating}
\usepackage{blindtext}
\usepackage{transparent}
\usepackage{setspace}
\usepackage[table]{xcolor}

\definecolor{maroon}{cmyk}{0,0.87,0.68,0.32}

\newcommand{\indep}{\mbox{$\,\perp\!\!\!\perp\,$}}
\newcommand{\nindep}{\mbox{$\,\not\!\perp\!\!\!\perp\,$}}
\newcommand{\q}{``}
\newcommand{\qq}{''\xspace}

\usetikzlibrary{shapes,arrows,shadows,decorations}
\usetikzlibrary{decorations.pathmorphing}
\usetikzlibrary{backgrounds}

\newcommand{\newpar}{{\vspace{0.15cm} \noindent}}

\begin{document}

\section*{\title{Mendelian Randomization with Incomplete Exposure Data: a Bayesian Approach}}
\title{Mendelian Randomization with Incomplete Exposure Data: a Bayesian Approach}

\author{Teresa Fazia$^{1,\dagger\ast}$,
Leonardo Egidi $^{2,\dagger}$,
Burcu Ayoglu $^{3}$,
Ashley Beecham $^{4}$,\\
Pier Paolo Bitti $^{5}$,
Anna Ticca $^{6}$,
Hui Guo $^{7}$,
Jacob L. McCauley $^{4}$,\\
Peter Nilsson $^{3}$,
Rosanna Asselta $^{8,9}$,
Carlo Berzuini $^{7\ddagger}$,
Luisa Bernardinelli $^{1\ddagger}$\\[4pt]
\rule{0cm}{1cm}
\textit{$^{\dagger}$} Joint First Authors,\\
\textit{$^{\ddagger}$} Joint Last Authors\\
\rule{0cm}{1cm}
\textit{$^{\ast}$} Corresponding Author: Teresa Fazia,
Department of Brain and Behavioral\\ Sciences,
Via Bassi 21. University of Pavia, 27100 Pavia, Italy.\\
e-mail: teresa.fazia01@ateneopv.it}

\date{}

% Running headers of paper:
\markboth
% First field is the short list of authors
{T.Fazia, L.Egidi and others}
% Second field is the short title of the paper
{Bayesian Mendelian Randomisation Analysis of Family Data}

\maketitle

\vspace{-0.7cm}
\noindent \textit{$^{1}$} Department of Brain and
Behavioural Sciences, University of Pavia, Pavia,
Italy \hfill\\
\textit{$^{2}$} Department of Economics, Business,
Mathematics and Statistics “Bruno de \hfill\\ Finetti” (DEAMS),
University of Trieste, Trieste, Italy \hfill\\
\textit{$^{3}$} SciLifeLab, Dept. of Protein Science,
KTH Royal Institute of Technology,\hfill\\
Stockholm, Sweden\\
\textit{$^{4}$} John P. Hussmann Institute for Human
Genomics and Dr John Macdonald\\ Foundation, Department
of Human Genetics, University of Miami, Miller\\
School of Medicine, Miami, USA\\
\textit{$^{5}$} Azienda Tutela Salute Sardegna.
ASSL Nuoro.
Immunoematologia e Medicina\\
Trasfusionale. Ospedale "San Francesco", Nuoro, Italy\\
\textit{$^{6}$} Azienda Tutela Salute Sardegna.
ASSL Nuoro. Neurologia e Stroke Unit.\\
Ospedale "San Francesco", Nuoro, Italy\\
\textit{$^{7}$} Centre
for Biostatistics,
The University of Manchester,
Jean McFarlane Building,\\
University Place, Oxford Road,
Manchester
M13 9PL, UK\\
\textit{$^{8}$} Department of Biomedical Sciences,
 Humanitas University, Pieve Emanuele, Italy;\\
\textit{$^{9}$} Humanitas Clinical and Research Center, 
Rozzano, Italy\\
\textit{$^{7}$} Centre
for Biostatistics,
The University of Manchester,
Jean McFarlane Building,\\
University Place, Oxford Road,
Manchester
M13 9PL, UK\\

% Add a footnote for the corresponding author if one
% has been identified in the author list
%\footnotetext{To whom correspondence should be addressed.}

\section*{Abstract}

\noindent We expand Mendelian Randomization (MR) methodology to deal with randomly missing data on either the exposure or the outcome variable, and furthermore with data from non-independent individuals (eg components of a family). Our method rests on the Bayesian MR framework proposed by Berzuini et al (2018), which we apply in a study of multiplex Multiple Sclerosis (MS) Sardinian families to characterise the role of certain plasma proteins in MS causation. The method is robust to presence of pleiotropic effects in an unknown number of instruments, and is able to incorporate inter-individual kinship information. Introduction of missing data allows us to overcome the bias introduced by the (reverse) effect of treatment (in MS cases) on level of protein. From a substantive point of view, our study results confirm recent suspicion that an increase in circulating IL12A and STAT4 protein levels does not cause an increase in MS risk, as originally believed, suggesting that these two proteins may not be suitable drug targets for MS.

\section*{Introduction}

Mendelian Randomization (MR)  \cite{Katan1986}  \cite{Voight2012} \cite{Burgess2015}
\cite{Berzuini2018c} principles are well accepted in the epidemiologic community. Once the appropriate assumptions are entertained, MR analysis can  be used to assess the effect of an "exposure" variable on the medical outcome of interest, even when the former is not experimentally controlled. In most MR studies so far, the exposure is a late actor in the biological cascade, for example obesity \cite{Conde2018}. Many such studies exploit the wealth of data gathered from follow-up of a large cohort of (initially) healthy population subjects.

\newpar This paper investigates methodological problems involved in the application of MR outside the above described standard scenario. We investigate problems encountered when the putative causal factor operates at a molecular (transcriptome or proteome) level (rather than deep down the biological cascade), whether with the aim of dissecting a disease pathway or discovering a new pharmacologic target. We also consider situations where the data are collected from a small/isolated population, where one can reap the advantages of multiplex family data analysis \cite{Brumpton2019}, and escape those biases in health research that are brought about by excessive reliance on western-world biobanks. The isolated population considered in the present paper belongs to a region of the Sardinian island, Italy.

\newpar We also consider situations where biases may arise due to {\em (i)}  the exposure in diseased subjects being possibly affected by treatment,  and {\em (ii)} random missingness of exposure values.

\newpar In this paper we advocate a Bayesian approach to MR in the above scenarios. We have constructed our proposed method out of the Bayesian MR framework described by Berzuini and colleagues \cite{Berzuini2018a}. A Bayesian approach to MR appears to suit the study scenarios of interest here. One reason being coherent handling of the uncertainty around the estimated instrument-exposure associations, as a safeguard against the weak-instrument bias \cite{Bound1995}. Another reason being freedom from asymptotic assumptions, as a safeguard against the limited size of the sample and the limited number of instrumental variants associated with a transcriptome/proteome-level exposure. And a final reason being coherent Bayesian handling of missing exposure values, which are treated as additional parameters to be estimated from the data, thereby avoiding the biases that affect two-sample frequentist analysis. Also illustrated here is the ease of eleboration of the Bayesian model to deal with inter-correlation of individuals within a family.

\newpar Our illustrative study addresses the question whether the concentration of certain inflammatory proteins,  IL12A in the first place, plays a causal role in the development of Multiple Sclerosis (MS). Current opinion holds that an increase in the circulating level of some of these proteins increases risk of MS. An MR analysis of our study data reverses this opinion. In particular, our analysis discredits existing hopes that an decrease in IL12A plasma concentration may reduce risk of the disease. This example also illustrates a change with respect to the prevailing MR analysis ethos. In the context of mechanistic studies like the one described in this paper, which sit at the frontier between epidemiology and biology, emphasis is not on the statistical significance of estimates, as much as on the use of data evidence to tip the balance between competing biological hypotheses. Fruitful interaction between biologists and statisticians often takes place in this sort of grey area.

\newpar The computational engine of the proposed method is the same as in  \cite{Berzuini2018a}, being based on recent developments  in Markov chain Monte Carlo (MCMC) inference \cite{Neal2011}, which have been implemented in the {\tt Stan} probabilistic programming  language \cite{carpenter2017}.

\newpar The "outcome" variable of our illustrative MR analysis is the  MS disease indicator. MS lends itself well to a MR study.  This disease tends to become manifest early during  reproductive lifespan of most humans, throughout  history, and is therefore likely to have a strong genetic  component. Genetic variants are therefore expected to act as good instruments for the MR analysis, although their number will be inevitably smaller than in the case of a high-level biomarker. The main  scientific question in our study is whether the plasma  level of IL12A protein is causal with respect to  development of MS. It is believed that  dysregulation of circulating proteins is a causal  determinant in many pathologies, more directly so  than genetic variants. Our analysis is further motivated  by the importance of proteins as natural drug targets.
\section*{Method}

\subsection*{Sample Description}

\noindent Our MS patients were ascertained through a case register established in 1995 in the province of Nuoro, Sardinia, Italy. Cases were diagnosed according to Poser’s criteria \cite{Poser1983}. Twenty extended MS multiplex pedigrees were selected for the analysis, for a total of $N=936$ individuals (98 cases and 838 unaffected relatives). A subset of the pedigree members had complete data, consisting of the observed levels of the IL12A protein (the exposure), the known disease indicator (the outcome variable), and the genotypes at all loci of  the Immunochip Illumina Infinium HD custom array (hereafter “Immunochip” for brevity), designed for fine mapping of 184 established autoimmune loci \cite{Beecham2013}. . The remaining individuals had complete data except for a missing value for the protein level.

\subsection*{Genotyping Data}

\newpar The quality control-filtered set of genotyping data included 127134 Single Nucleotide Polymorphisms (SNPs) across Immunochip
\cite{Fazia2017}. We imposed a maximum correlation of $r^2 = 0.20$ between candidate instrumental SNPs within a 100 Kb window, by using the {\tt indep-pairwise} command of the PLINK package \cite{Purcell2007}. This yielded a total of 19121 candidate SNP instruments across Immunochip.

\subsection*{Protein Selection and Profiling}

\noindent The proteins we chose for our illustrative study were IL12A, IL23A, IL4, STAT4, and STAT6.
Choice was made prior to considering the data, on the basis of Genome-Wide Significant (GWS) association between MS and genetic variants located within (e.g. exonic, intronic, in the UTR) or in the proximity (e.g. downstream, intergenic) of the protein-coding gene \cite{Beecham2013}, their  involvemente in immune response and the availability in our sample of data. 
\newpar Plasma profiles were analysed by using a bead-based antibody array format, consisting of polyclonal Human Protein Atlas
\cite{Nilsson2005} 
 antibodies immobilized onto microspheres in suspension
\cite{Schwenk2007} \cite{Schwenk2008} (see Supplementary Material for details).

\subsection*{Selection of Instrumental Variants}
\noindent Genetic variants with a significant marginal association ($p<5 \times 10^{-3}$) with the level of the protein of interest and mutual  $r^2<0.20$ correlation were selected to act as instrumental variables (IVs) in the first stage of our analysis. The liberal $p<5 \times 10^{-3}$ threshold is justified by the fine genotyping of candidate gene regions and by recent arguments
\cite{Wang2016} \cite{Yang2011}
in favour of using sub-genome-wide-significance loci to strengthen biologically interesting signals. It is also justified by the relative ability of our Bayesian MR method (when compared with most frequentist approaches) to deal with the weak instrument bias, thanks to the uncertainty of the estimated exposure coefficients being explicitly included in the model.

\subsection*{Notation}

   \noindent Let, for example, the circulating level of protein IL12A act as the exposure in the analysis, and be denoted as $X$. Let the symbol $\Sigma_X$  denote a regime indicator \cite{Dawid2000} \cite{Dawid2002}, with  $\Sigma_X = \emptyset$ telling us that $X$ is to be interpreted as the result of a passive observation (observational regime), while  $\Sigma_X = x$ tells us that the value if $X$ is to be interpreted as fixed under a {\em hypothetical} interventional regime to be equal to $x$. The outcome variable, $Y$, indicates whether the individual has the disease ($Y=1$) or not ($Y=0$). We are interested in the "causal effect" of $X$ on $Y$, that is, in the way the distribution of $Y$ changes when $X$ is first set by intervention to a reference value $x_0$ and then forced to take the new value $x_1$. Throughout this paper we take this causal effect to be defined as the causal odds ratio (COR):
\begin{equation}
COR = \frac{P(Y=1\mid \Sigma_X = x_1)} {P(Y=1\mid \Sigma_X=x_0)}   \frac{1-P(Y=1\mid \Sigma_X = x_0)} {1-P(Y=1\mid \Sigma_X=x_1)} 
\end{equation}
\noindent  Presence of the unobserved confounders of the exposure-outcome relationship, denoted as $U$, is the reason why we can't generally measure the causal effect of $X$ on $Y$ by standard regression of $Y$ on $X$. A causally meaningful estimate of  that effect can, under assumptions, be obtained via MR. We shall model $U$ as an individual-level scalar variable, more precisely, a one-dimensional reduction of the unknown collection of confounders. MR requires availability of a set of instrumental variables,  or instruments, denoted as $Z  \equiv (Z_1, \ldots , Z_J) $, which in a standard analysis will often correspond to the individual's genotypes at a set of SNP loci. Each of these genotypes we code as "allele doses", with values $(0,1,2)$ respectively indicating presence of zero, one and two copies of the "alternative" allele at the locus. For most individuals in the pedigree, we also have observed {\em (i)} maternal and paternal genotypes at each instrumental locus and {\em (ii)} the levels of protein IL12A in the father and in the mother. We further introduce an individual-level categorical variable, denoted as $F$, which indicates the individual's pedigree of membership, or family. Further notation will be introduced in the next sections, as required.

\subsection*{Assumptions}

\noindent  We are going to describe conditions for validity of the method by
using Dawid's conditional independence formalism
\cite{Dawid1979}, with the $A \indep B \mid C$ notation
standing 
for \q$A$ is independent of $B$ given $C$, and $A \nindep B$, meaning \q$A$ 
is not independent of $B$ \qq.  The conditions
we are going to introduce  are essentially
identical to those required by standard MR  methods.

\newpar Each $j$th instrumental variable, $Z_j$,
must satisfy the {\em confounder independence} condition $Z_j \indep U$, 
stating that the instrument is unrelated to exposure-outcome confounders. 
This condition is not directly testable.
A further condition called {\em exclusion-restriction} requires that
 $Y \indep Z_j \mid (X,U)$, that is,  each $j$th instrument can be 
associated with response only via the exposure.
This condition cannot be directly tested.
Exclusion-restriction 
is a desirable condition, however, unlike the remaining conditions in 
this section, it is not required by our method. Next comes the {\em 
instrument relevance} condition, $Z_j \nindep X$, stating that no 
instrument is independent of the exposure. We have also conditions 
involving the regime indicator, $\Sigma_X$.  The {\em confounder 
invariance} condition, $U \indep \Sigma_X$, requires that the 
distribution of the confounders $U$ be the same, whether or not we 
intervene on $X$, and regardless of the value imposed on or observed 
in $X$. Next comes the {\em interventional irrelevance} condition
$\Sigma_X \indep Z$, requiring that any intervention on $X$ has no 
consequence on $Z$, and the {\em interventional modularity} 
condition, $\Sigma_X \indep Y \mid (X,U)$, asserting that once 
we are told the values of $X$ and $U$, the distribution of $Y$ no 
longer depends on the way the value of $X$ has arisen, whether 
observationally or through the intervention of interest.

\newpar All the conditions defined above, except for exclusion-restriction, 
are required by our method.

\newpar Sometimes it is possible, and then helpful, to represent
the qualitative structure of a statistical model by a directed acyclic 
graph \cite{Lauritzen1996}. A stripped-down representation of the 
class of MR models discussed in the present paper
is shown in Figure 1. In this geaph, the variable denoted by the symbol $F$
is a family indicator, which we represent in the model equations
as a categorical factor with one level per family. Had we not allowed
for the presence of this indicator, we would have an unblocked path $Z \leftarrow F \rightarrow Y$ path (in a $d$-separation terminology) corresponding to a spurious, exposure-unmediated, association  between instrument and outcome. This would violate the Confounder Independence assumption, and consequently lead to a biased estimate of the causal effect.  In situations where the sample contains unrelated (in  addition to related) individuals, the unrelateds may be  lumped into a single, notional, family.

\newpar All the
assumptions/conditions stated at the beginning of this section
 (except for exclusion-restriction)
can be read off the graph of Figure 1 by applying 
$d$-separation \cite{Geiger1990} or moralization \cite{Lauritzen1996},
with the following additional rules:  {\em (i)} faithfulness \cite{Spirtes2001} 
of the $Z \rightarrow X$  edges ), meaning no instrument is independent of the exposure,
and  {\em (ii)} assigning a value $x$ to $\Sigma_X$ implies 
the simultaneous assignment of the same value to $X$, and  {\em (iii)}  
assigning a value $x$ to $\Sigma_X$ implies that all arrows into $X$ 
except for $\Sigma_X \rightarrow X$ are severed. Because most of the 
conditions introduced at the beginning of this section are not directly 
testable on the basis of the data, the reader should be aware that 
graphs like the one shown in Figure 1 describe an {\em assumed}, 
ultimately uncertified, albeit plausible, state of affairs. We shall assume 
throughout the paper that the above described conditions, bar 
exclusion-restriction, are valid. 

\newpar We conclude this section with a brief discussion of the 
exclusion-restriction assumption. This assumption (which is not 
required by our method) does not allow an instrument to exert 
an effect on $Y$ other than that exerted though the mediating 
effect of $X$. In our graph of Figure 1, this condition is violated 
by the $Z_J \rightarrow Y$ arrow. Because of this, the effect of 
instrument $Z_J$ on $Y$ is said to be \q pleiotropic \qq according 
to Figure 1. In the context of our application, pleiotropic effects 
may arise from two broad classes of mechanism. The first is due 
to the variants used as instruments being in linkage 
disequilibrium (LD) with variants of nearby genes. The second is 
due to the instrumental variant exerting a causal effect on $Y$ 
through a pathway independent of $X$. Although the former type 
of pleiotropy could, in principle, be neutralized by conditioning on 
the variants  in the region, except for the instrumental variants, the 
latter cannot be directly tested from the data. It would therefore be 
uncautious to perform MR by using a method that does {\em not} 
allow for general types of pleiotropy. Our Bayesian approach deals 
with the problem by explicitly introducing the unknown pleiotropic 
effects in the model, and by treating them as unknown parameters 
to be estimated from the data.

%Figura 1. Grafo modello di indipendenza con variabile famigliare.
\begin{center}
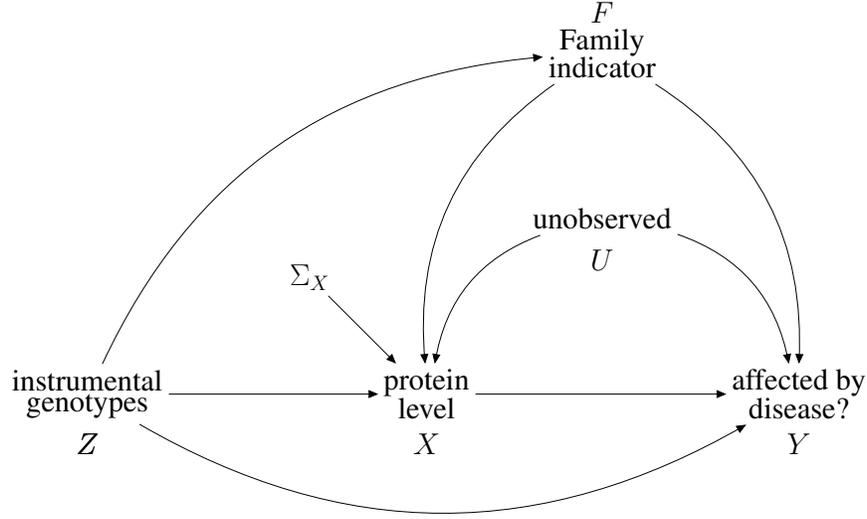
\begin{figure}
\centering
\caption{\footnotesize Graphical representation of the model.}
\vspace{1cm}
\scalebox{0.55}{
\begin{tikzpicture}[align=center,node distance=4.5cm]
\node (Uchild) [fill=none] {\LARGE unobserved};
\node (Ulab) [below of=Uchild,fill=none,node distance=1cm]
{\LARGE $U$};
\node (family) [above of=Uchild, fill=none,node distance=4cm]
{\LARGE Family\\ \LARGE indicator};
\node (Flab) [above of=family,fill=none,node distance=1cm]
{\LARGE $F$};
\node (Xchild) [below left of=Uchild,fill=none,rectangle,node distance=6cm]
{\LARGE  \LARGE protein \\ \LARGE level};
\node (Xlabchild) [below of=Xchild,fill=none,node distance=1.2cm]
{\LARGE $X$};
\node (Ychild) [right of=Xchild,fill=none,node distance=9cm]
{\LARGE affected by\\ \LARGE disease?};
\node (Ylabchild) [below of=Ychild,fill=none,node distance=1
.2cm] {\LARGE $Y$};
\node (handle) [left of=Xchild,fill=none,node distance=3.7cm]{};
\node (Zchild) [left of=handle,fill=none,rectangle]
{\LARGE  \LARGE instrumental \\ \LARGE genotypes};
\node (Zlabchild) [below of=Zchild,fill=none,node distance=1.2cm] {\LARGE $Z$};
\node (Zlabchild) [below of=Zchild,fill=none,node distance=1.2cm] {\LARGE $Z$};
\node (F) [above left of=Xchild,fill=none,node distance=4cm]
{\LARGE $\Sigma_X$};
\draw[-triangle 45] (F) to node {} (Xchild);
\draw[-triangle 45, bend left] (Zchild) to node {} (family);
\draw[-triangle 45,bend left] (Uchild) to node {} (Ychild);
\draw[-triangle 45, bend right] (Uchild) to node {} (Xchild);
\draw[-triangle 45] (Zchild) to node {} (Xchild);
\draw[-triangle 45] (Xchild) to node {} (Ychild);
%\draw[-triangle 45] (family) to node {} (Uchild);
\draw[-triangle 45,bend right] (family) to node {} (Xchild);
\draw[-triangle 45, bend left] (family) to node {} (Ychild);
\draw[-triangle 45, bend right] (Zchild) to node [above] {} (Ychild);
\end{tikzpicture}
}
\end{figure}
\end{center}

\subsection*{Quantitative Specification of the Model}

\noindent The graphical representation in Figure 1  is an elaboration
of the model of Berzuini and colleagues
\cite{Berzuini2018a}. We note that in our representation the unobserved variable $U$ is explicitly included in the model, rather than integrated out. This choice does not affect the likelihood function but will generally lead to different prior distributions. In the following equations the continuous exposure variable,  $X$, is implicitly assumed  to be transformed to a zero mean and unit standard deviation variable. In this preliminary version of the model,  the individuals are treated as conditionally independent given the family indicator, $F$. The model allows for an effect of the family on both the exposure and the outcome, but assumes that family and exposure do not interact in their effects on the outcome. The Reader may wish to check that the graph (and the model equations) conform to the conditional independence assumptions of the preceding section.
\begin{eqnarray*}
\label{full1}
P(U) &=& \mbox{N}(0,1),\\
\label{full2}
\label{family1}
P(X \mid Z_1, \ldots , Z_J, U, F) &=& \mbox{N}(\nu,\sigma_X^2),\\
\nu &=& \sum_{j=1}^J \alpha_j Z_j +
%\gamma_{j X}_g(P) +
\delta_X U+
\sum_{f=1}^{M} I_{F=f} \; \gamma^X_f ,\\
\label{family2}
P(Y \mid X, Z_1, \ldots , Z_J, U, F) &=& {\rm Bernoulli}(\pi),\\
logit(\pi) &=& \mbox{MVN}(\mu, \Sigma),\\
\mu &=& \omega_Y + \theta X +
\sum_{j=1}^J \beta_j Z_j + U+
\sum_{f=1}^{M} I_{F=f}\; \gamma^Y_f,
\end{eqnarray*}
\noindent where $\mbox{N}(a,b)$ stands for a
normal distribution with
mean $a$ and variance $b$, the symbol
$\alpha
\equiv (\alpha_1, \ldots , \alpha_J)$
denotes the instrument-exposure associations, and
$\beta
\equiv (\beta_1, \ldots , \beta_J)$
are the pleiotropic effects. The notation $I_A$
stands for the indicator function, taking 
value 1 if the logical condition $A$ is true, and value 0 otherwise. 
The quantities $\gamma^X \equiv (\gamma^X_1, \ldots , 
\gamma^X_M)$ and $\gamma^Y \equiv (\gamma^Y_1, 
\ldots , \gamma^Y_M)$ are vectors of unknown "family 
effects" acting upon $U$. In our analysis, we 
have imposed on these effects independent and mildly 
informative priors, with greater spread than the prior 
for $\theta$.

\newpar Unlike the model by Berzuini et al, the outcome variable $Y$ here is distributed as a Bernoulli random variable, as appropriate for a binary outcome. The causal effect of interest, denoted as $\theta$, represents the change in log-odds of probability of $Y=1$ caused by an interventional change of one standard deviation in $X$. Family information is incorporated by designating a categorical variable $F$ to indicate the individual's family, with $F \in (1, \ldots, M)$, and  $M$ denoting the total number of families.

\newpar Recall that, in our study, some components of the $X$ vector (protein level measurements) are missing, which is not made explicit in the notation above. The Bayesian inference engine identifies the missing components and treats them as unknown parameters, effectively integrating them out to obtain the posterior distribution for the parameters of inferential interest. Note that this way of dealing with missing data is more efficient than, say, imputing each missing component of $X$ on the basis of the individual's observed $Z$ values, thanks to the fact that, in our method, the missing values are estimated by using information about both $X$ and $Y$.

\newpar As shown in \cite{Berzuini2018a} for the normal case,
parameters  $(\alpha,
\tau_X)$ are identified by the data, but
the remaining parameters, including
the causal effect of interest, $\theta$,
are not. Berzuini  and colleagues deal with the problem by a combination of two devices. The first consists of introducing the additional (untestable) assumption that each $j$th component of $\beta$ is a priori independent of the remaining parameters of the model, formally,
$P(\beta_j \mid \alpha_j,\tau_X) = P(\beta_j)$.
This is called the Instrument Effects Orthogonality (IEO) condition.
The second consists of introducing
a proper, scientifically plausible, prior for $\beta$, which makes inferences possible by inducing on $\theta$ (and on further parameters of potential posterior interest) a proper posterior. 

\noindent As concerns the prior component of our Bayesian model, we invite the Reader to consult \cite{Berzuini2018a}.

\newpar Variations from \cite{Berzuini2018a} have been introduced. While still imposing on the pleiotropic effects $\beta$ a horseshoe prior \cite{carvalho2010}, we are now using the enhanced version of this distribution proposed by Piironen and Vehtari \cite{Piironen2017}. Also, we take $\theta$ -- the causal effect of main inferential interest -- to have a Cauchy(0,2.5) prior, with the following justification. Because $X$ has been standardized to have mean 0 and unit standard deviation (SD), the mentioned prior for $\theta$ states as unlikely that a one-SD change in protein level causes a change in risk of disease exceeding 5 points on a logit scale, which corresponds to shifting a probability of disease  occurrence from, say, 0.01, to 0.5, or from 0.5 to 0.99. This is also in agreement with current evidence on the effect of circulating proteins on disease  \cite{Sun2018}.

\newpar Finally, we are now taking the instrument-exposure associations, $\alpha$, to be independently distributed according to a double-exponential distribution with mean 0 and unknown scale. One merit of this prior is to shrink the small effects to zero, which reduces the weak instrument bias, so that the model works with an adaptively selected subset of strong instruments.

\subsection*{Introducing Kinship}

\noindent The preceding model treats members of a pedigree as independent individuals, which they are not. This will produce overconfident and biased estimates. We remedy this by introducing in the model between-individual correlation in the form of the kinship matrix, which can be derived by a standard algorithm from the structure of the pedigree. We are currently working with a single, overarching, kinship matrix of size $N \times N$, where $N$ is the total number of individuals in the sample. This large matrix contains zeros corresponding to pairs of individuals in different families. The method could be made computationally more efficient by introducing  family-specific matrices. Kinship information is introduced in the model by replacing the previous specification for $Y$ with:

\begin{eqnarray*}
\label{kinship1}
\label{full3}
P(Y \mid X, Z_1, \ldots , Z_J, U, F) &=& {\rm Bernoulli}(\pi),\\
logit(\pi) &=& \mbox{MVN}(\mu, \Sigma),\\
\mu &=& \omega_Y + \theta X +
\sum_{j=1}^J \beta_j Z_j + U+
\sum_{f=1}^{M} I_{F=f}\; \gamma^Y_f,
\end{eqnarray*}
\noindent where $\Sigma$ is the $N \times N$ kinship matrix,
the notation $\mbox{MVN}   (a,b)$ stands for multivariate normal distribution with vector mean $a$ and variance-covariance matrix $b$.

\section*{Analysis Strategy and Results}

\subsection*{Frequentist Analysis}

\vspace{0.4cm}

\noindent We have separately studied the five proteins of interest via a standard battery of frequentist MR algorithms, as offered by {\tt R} package {\tt MendelianRandomization} \cite{yavorska2017}, as found on  http://cran.r-project.org. 

\newpar For each individual protein, we proceeded by fitting a linear mixed-effects regression model of the dependence of the circulating protein level, $X$, on each separate instrumental SNP, adjusting for sex, with familial relatedness between individuals accounted for in the model through the kinship matrix \cite{Pinheiro2000}, by using the {\tt lmekin} R function. Each of these regressions, of $X$ on $Z_j$ say, yields an estimated coefficient $\Phi_j$ and a corresponding standard error. Only those disease-free individuals with a measured value of the $X$ were involved in these regressions.  We then performed on the entire sample a logit-linear, sex-adjusted, regression of $Y$ on each $j$th SNP, and let the resulting effect estimate, on a log-odds ratio scale, be denoted as $\Gamma_j$. The estimates of the $\{\Phi_j\}$ and of the $\{\Gamma_j\}$ acted as an input to our frequentist MR estimate of the COR effect of $X$ on $Y$, and this analysis was replicated by using each of the following frequentist MR methods: Inverse-Variance Weighted estimator (IVW) \cite{Bowden2015} , the MR-Egger Regression estimator \cite{Bowden2016}  and Weighted Median Estimator \cite{Bowden2015}. These methods, as provided by the mentioned  {\tt MendelianRandomization} R package, are able to work from the  $\Phi_j$ and $\Gamma_j$  statistics to obtain the estimate of the causal effect of interest, and assume the instruments to be independent.

\newpar The above approach acknowledges that exposure values in the diseased  may have been reverse-affected by treatment, and wisely discards them from the analysis, as if they were missing. This, however, incurs bias due to using control data twice.

\newpar Results of these analyses are summarized below by plotting marginal effects of each instrument on log-odds risk of disease against the corresponding effects on exposure (each of the "Egger plots" consists of a scatter diagram where each instrument is represented as a dot, with horizontal coordinate  $\{\Phi_j\}$ and vertical coirdinate $\{\Gamma_j\}$ ), and, in addition, in Table \ref{Frequentist}, by reporting the estimated causal effects, for each protein and algorithm, together with their standard errors, 95\% confidence intervals and P-values. Estimated intercepts from the Egger methods are not reported.

\newcommand{\ra}[1]{\renewcommand{\arraystretch}{#1}}

\newcommand{\midheader}[2]{%
      %  \midrule
\topmidheader{#1}{#2}}
\newcommand\topmidheader[2]{\multicolumn{#1}{c}{\textsc{#2}}\\%
                \addlinespace[0.5ex]}

\vspace{1.5cm}

%\begin{table*}[ht]
%\centering
%\ra{1.3}
\ra{1}
%\begin{longtable}{llccccc}
\begin{longtable}{ll*5{p{1.5cm}}}
\caption{\small Results from a frequentist analysis of the causal effects of 
five proteins on susceptibility to MS. The second column
of the table indicates the
MR algorithm.\label{Frequentist}}\\
\rowcolor{maroon!10}
\topmidheader{2}{PROTEIN IL4}
%&& \multicolumn{5}{c}{CAUSAL EFFECT OF PROTEIN ON MS}\\
%\cmidrule{3-7}
\rowcolor{white}
&& Estimate & Std Error & \multicolumn{2}{c}{95\% CI}&  P-value \\ 
\cmidrule{5-6}
\endfirsthead
\caption[]{(continued)}\\
%&& \multicolumn{5}{c}{CAUSAL EFFECT OF PROTEIN ON MS}\\
%\cmidrule{3-7}
\rowcolor{white}
%\vspace{0.2cm}
&& Mean & Std Error & \multicolumn{2}{c}{95\% CI}&  P-value \\ 
\cmidrule{5-6}
\endhead
%\toprule\\
 1 & Simple median & 0.07 & 0.07 & -0.06 & 0.21 & 0.28 \\ 
2 & Weighted median & 0.09 & 0.07 & -0.05 & 0.22 & 0.21 \\ 
3 & Penalized weighted median & 0.08 & 0.07 & -0.05 & 0.22 & 0.21 \\ 
4 & IVW & 0.09 & 0.05 & -0.00 & 0.18 & 0.05 \\ 
5 & Penalized IVW & 0.08 & 0.05 & -0.01 & 0.18 & 0.08 \\ 
6 & Robust IVW & 0.08 & 0.04 & -0.00 & 0.17 & 0.06 \\ 
7 & Penalized robust IVW & 0.08 & 0.04 & -0.00 & 0.17 & 0.06 \\ 
8 & MR-Egger & 0.03 & 0.33 & -0.62 & 0.67 & 0.94 \\ 
%9 & (intercept) & 0.02 & 0.10 & -0.18 & 0.22 & 0.84 \\ 
9& Penalized MR-Egger & 0.00 & 0.33 & -0.64 & 0.65 & 0.99 \\ 
%11 & (intercept) & 0.03 & 0.10 & -0.18 & 0.23 & 0.80 \\ 
10& Robust MR-Egger & -0.02 & 0.27 & -0.56 & 0.51 & 0.93 \\ 
%13 & (intercept) & 0.03 & 0.09 & -0.14 & 0.20 & 0.69 \\ 
11& Penalized robust MR-Egger& -0.03 & 0.27 & -0.56 & 0.51 & 0.92 \\ 
%15 & (intercept) & 0.03 & 0.09 & -0.13 & 0.20 & 0.69 \\ 
\\
%questo sopra fatto
\rowcolor{maroon!10}
\topmidheader{2}{PROTEIN IL4}
\\
  1 & Simple median & 0.07 & 0.07 & -0.07 & 0.21 & 0.31 \\ 
2 & Weighted median & 0.09 & 0.07 & -0.05 & 0.22 & 0.21 \\ 
3 & Penalized weighted median & 0.09 & 0.07 & -0.05 & 0.22 & 0.21 \\ 
4 & IVW & 0.09 & 0.05 & 0.00 & 0.19 & 0.05 \\ 
5 & Penalized IVW & 0.09 & 0.05 & -0.01 & 0.18 & 0.07 \\ 
6 & Robust IVW & 0.09 & 0.04 & -0.00 & 0.17 & 0.05 \\ 
7 & Penalized robust IVW & 0.08 & 0.04 & -0.00 & 0.17 & 0.05 \\ 
8 & MR-Egger & 0.14 & 0.34 & -0.53 & 0.81 & 0.68 \\ 
%9 & (intercept) & -0.01 & 0.11 & -0.22 & 0.19 & 0.89 \\ 
9& Penalized MR-Egger & 0.12 & 0.34 & -0.55 & 0.79 & 0.73 \\ 
%11 & (intercept) & -0.01 & 0.11 & -0.22 & 0.20 & 0.93 \\ 
10& Robust MR-Egger & 0.09 & 0.25 & -0.41 & 0.59 & 0.72 \\ 
%13 & (intercept) & -0.00 & 0.08 & -0.16 & 0.16 & 0.98 \\ 
11& Penalized robust MR-Egger & 0.09 & 0.26 & -0.41 & 0.59 & 0.73 \\ 
%15 & (intercept) & -0.00 & 0.08 & -0.16 & 0.16 & 0.99 \\
\\
%questo sopra fatto
\rowcolor{maroon!10}
\topmidheader{2}{PROTEIN IL23A}
\\
1 & Simple median & -0.02 & 0.06 & -0.14 & 0.09 & 0.67 \\ 
2 & Weighted median & 0.04 & 0.06 & -0.07 & 0.15 & 0.51 \\ 
3 & Penalized weighted median & 0.04 & 0.06 & -0.07 & 0.15 & 0.51 \\ 
4 & IVW & 0.04 & 0.04 & -0.04 & 0.12 & 0.31 \\ 
5 & Penalized IVW & 0.04 & 0.04 & -0.04 & 0.12 & 0.31 \\ 
6 & Robust IVW & 0.03 & 0.05 & -0.06 & 0.13 & 0.53 \\ 
7 & Penalized robust IVW & 0.03 & 0.05 & -0.06 & 0.13 & 0.53 \\ 
8 & MR-Egger & 0.40 & 0.19 & 0.01 & 0.78 & 0.04 \\ 
%9 & (intercept) & -0.10 & 0.06 & -0.21 & 0.01 & 0.06 \\ 
9& Penalized MR-Egger & 0.40 & 0.19 & 0.01 & 0.78 & 0.04 \\ 
%11& (intercept) & -0.10 & 0.06 & -0.21 & 0.01 & 0.06 \\ 
10& Robust MR-Egger & 0.43 & 0.27 & -0.09 & 0.95 & 0.11 \\ 
%13 & (intercept) & -0.11 & 0.07 & -0.26 & 0.03 & 0.11 \\ 
11& Penalized robust MR-Egger & 0.43 & 0.27 & -0.09 & 0.95 & 0.11 \\ 
%15 & (intercept) & -0.11 & 0.07 & -0.26 & 0.03 & 0.11 \\ 
\\
%questo sopra fatto
\rowcolor{maroon!10}
\topmidheader{2}{PROTEIN IL12A}
\\
 1 & Simple median & -0.17 & 0.10 & -0.36 & 0.02 & 0.08 \\ 
2 & Weighted median & -0.12 & 0.09 & -0.31 & 0.06 & 0.19 \\ 
3 & Penalized weighted median & -0.13 & 0.09 & -0.31 & 0.06 & 0.18 \\ 
4 & IVW & -0.11 & 0.07 & -0.24 & 0.02 & 0.09 \\ 
5 & Penalized IVW & -0.11 & 0.07 & -0.24 & 0.02 & 0.09 \\ 
6 & Robust IVW & -0.12 & 0.06 & -0.24 & -0.00 & 0.05 \\ 
7 & Penalized robust IVW & -0.12 & 0.06 & -0.24 & -0.00 & 0.05 \\ 
8 & MR-Egger & -0.05 & 0.25 & -0.53 & 0.44 & 0.85 \\ 
%9 & (intercept) & -0.02 & 0.06 & -0.14 & 0.10 & 0.78 \\ 
9& Penalized MR-Egger & -0.05 & 0.25 & -0.53 & 0.44 & 0.85 \\ 
%11 & (intercept) & -0.02 & 0.06 & -0.14 & 0.10 & 0.78 \\ 
10& Robust MR-Egger & -0.04 & 0.22 & -0.47 & 0.40 & 0.87 \\ 
%13 & (intercept) & -0.02 & 0.06 & -0.14 & 0.10 & 0.72 \\ 
11& Penalized robust MR-Egger & -0.04 & 0.22 & -0.47 & 0.40 & 0.87 \\ 
%15 & (intercept) & -0.02 & 0.06 & -0.14 & 0.10 & 0.72 \\ 
\\
%questo sopra fatto
\rowcolor{maroon!10}
\topmidheader{2}{PROTEIN IL12A}
\\
1 & Simple median & -0.16 & 0.10 & -0.35 & 0.03 & 0.09 \\ 
2 & Weighted median & -0.12 & 0.09 & -0.31 & 0.06 & 0.19 \\ 
3 & Penalized weighted median & -0.13 & 0.09 & -0.31 & 0.06 & 0.18 \\ 
4 & IVW & -0.11 & 0.07 & -0.24 & 0.02 & 0.09 \\ 
5 & Penalized IVW & -0.11 & 0.07 & -0.24 & 0.02 & 0.09 \\ 
6 & Robust IVW & -0.12 & 0.06 & -0.23 & -0.00 & 0.05 \\ 
7 & Penalized robust IVW & -0.12 & 0.06 & -0.23 & -0.00 & 0.05 \\ 
8 & MR-Egger & -0.05 & 0.25 & -0.53 & 0.44 & 0.85 \\ 
%9 & (intercept) & -0.02 & 0.06 & -0.14 & 0.10 & 0.78 \\ 
9& Penalized MR-Egger & -0.05 & 0.25 & -0.53 & 0.44 & 0.85 \\ 
%11 & (intercept) & -0.02 & 0.06 & -0.14 & 0.10 & 0.78 \\ 
10& Robust MR-Egger & -0.04 & 0.22 & -0.46 & 0.39 & 0.87 \\ 
%13 & (intercept) & -0.02 & 0.06 & -0.14 & 0.09 & 0.71 \\ 
11& Penalized robust MR-Egger & -0.04 & 0.22 & -0.46 & 0.39 & 0.87 \\ 
%15 & (intercept) & -0.02 & 0.06 & -0.14 & 0.09 & 0.71 \\ 
\\
%questo sopra fatto
\rowcolor{maroon!10}
\topmidheader{2}{PROTEIN STAT4}
\\
1 & Simple median & -0.02 & 0.08 & -0.17 & 0.13 & 0.80 \\ 
2 & Weighted median & -0.07 & 0.07 & -0.21 & 0.08 & 0.37 \\ 
3 & Penalized weighted median & -0.07 & 0.07 & -0.21 & 0.08 & 0.37 \\ 
4 & IVW & -0.09 & 0.05 & -0.19 & 0.01 & 0.08 \\ 
5 & Penalized IVW & -0.09 & 0.05 & -0.19 & 0.01 & 0.08 \\ 
6 & Robust IVW & -0.07 & 0.06 & -0.19 & 0.04 & 0.18 \\ 
7 & Penalized robust IVW & -0.07 & 0.06 & -0.19 & 0.04 & 0.18 \\ 
8 & MR-Egger & -0.03 & 0.17 & -0.37 & 0.31 & 0.87 \\ 
%9 & (intercept) & -0.02 & 0.05 & -0.12 & 0.08 & 0.71 \\ 
9& Penalized MR-Egger & -0.03 & 0.17 & -0.37 & 0.31 & 0.87 \\ 
%11 & (intercept) & -0.02 & 0.05 & -0.12 & 0.08 & 0.71 \\ 
10& Robust MR-Egger & -0.10 & 0.14 & -0.37 & 0.17 & 0.47 \\ 
%13 & (intercept) & 0.01 & 0.05 & -0.09 & 0.10 & 0.86 \\ 
11& Penalized robust MR-Egger & -0.10 & 0.14 & -0.37 & 0.17 & 0.47 \\ 
%15 & (intercept) & 0.01 & 0.05 & -0.09 & 0.10 & 0.86 \\ 
\\
%questo sopra fatto
\rowcolor{maroon!10}
\topmidheader{2}{PROTEIN STAT6}
\\
1 & Simple median & -0.06 & 0.06 & -0.18 & 0.05 & 0.25 \\ 
2 & Weighted median & 0.02 & 0.05 & -0.08 & 0.13 & 0.67 \\ 
3 & Penalized weighted median & 0.03 & 0.05 & -0.07 & 0.14 & 0.56 \\ 
4 & IVW & -0.04 & 0.04 & -0.11 & 0.03 & 0.22 \\ 
5 & Penalized IVW & -0.04 & 0.04 & -0.11 & 0.03 & 0.22 \\ 
6 & Robust IVW & -0.04 & 0.04 & -0.11 & 0.03 & 0.28 \\ 
7 & Penalized robust IVW & -0.04 & 0.04 & -0.11 & 0.03 & 0.28 \\ 
8 & MR-Egger & 0.15 & 0.09 & -0.03 & 0.34 & 0.10 \\ 
%9 & (intercept) & -0.08 & 0.04 & -0.15 & -0.01 & 0.02 \\ 
9& Penalized MR-Egger & 0.15 & 0.09 & -0.03 & 0.34 & 0.10 \\ 
%11 & (intercept) & -0.08 & 0.04 & -0.15 & -0.01 & 0.02 \\ 
10& Robust MR-Egger & 0.16 & 0.07 & 0.02 & 0.29 & 0.03 \\ 
% 13& (intercept) & -0.08 & 0.03 & -0.14 & -0.02 & 0.01 \\ 
11& Penalized robust MR-Egger & 0.16 & 0.07 & 0.02 & 0.29 & 0.03 \\ 
%15 & (intercept) & -0.08 & 0.03 & -0.14 & -0.02 & 0.01 \\ 
\end{longtable}
%\end{table*}

\begin{figure}[h!]
\vspace{-0.6cm}
\begin{subfigure}{0.9\textwidth}
\includegraphics[width=0.5\textwidth,height=5cm]{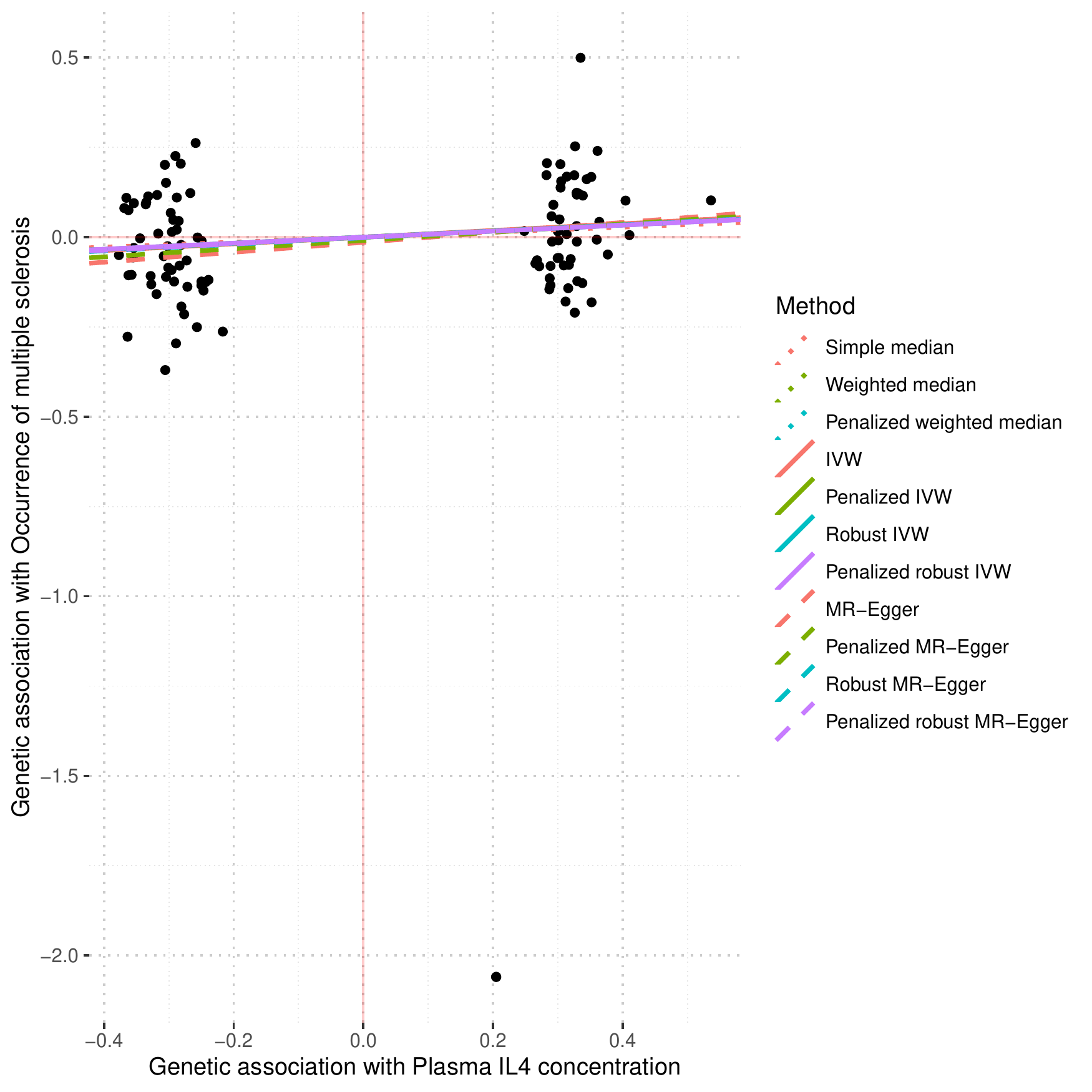}\hspace{1.5cm}
\includegraphics[width=0.5\textwidth,height=5cm]{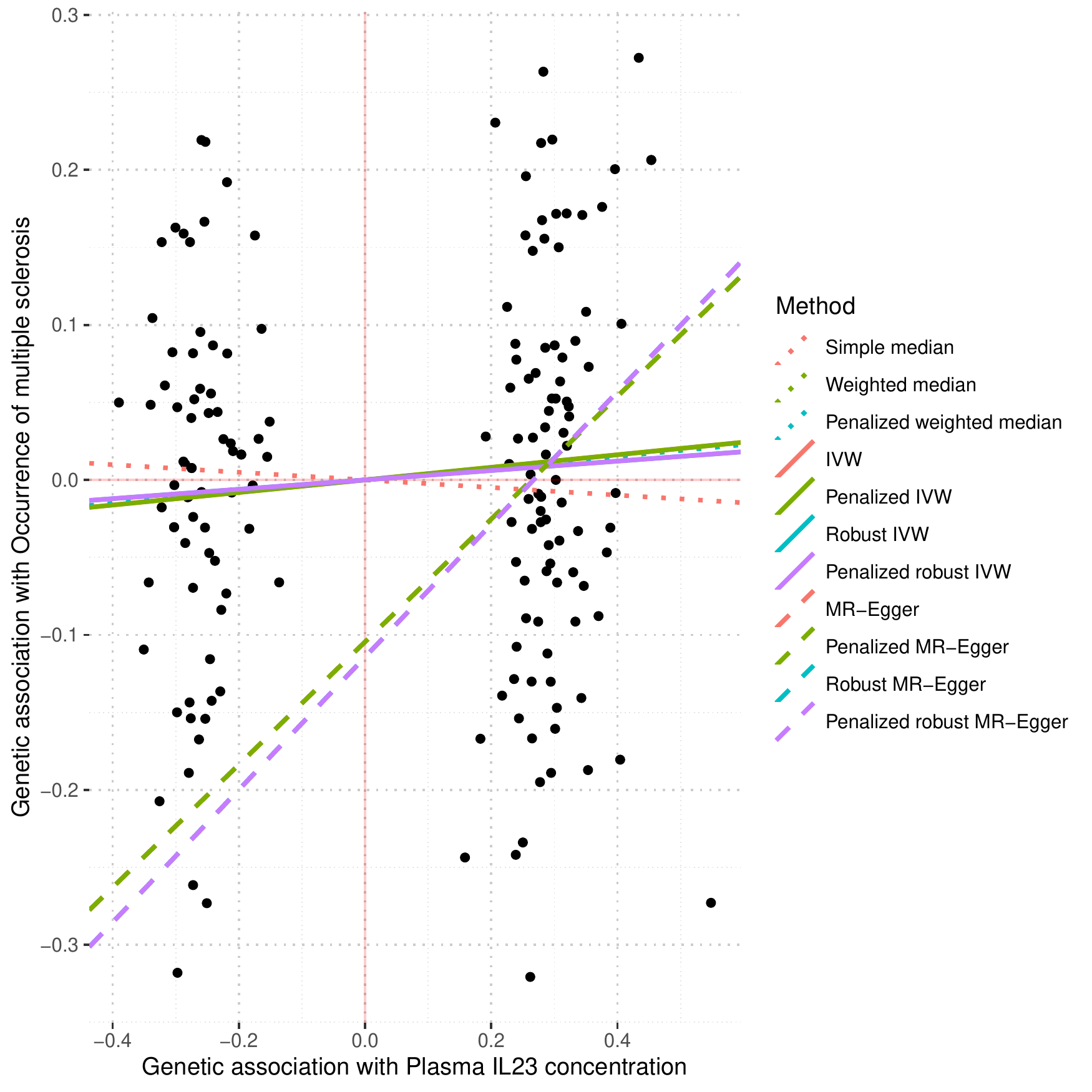}
\label{EggerPlotIL4IL23A}
\caption{\footnotesize Egger plots for proteins IL4 and IL23}
\end{subfigure}

\vspace{0.2cm}

\begin{subfigure}{0.9\textwidth}
\includegraphics[width=0.5\textwidth,height=5cm]{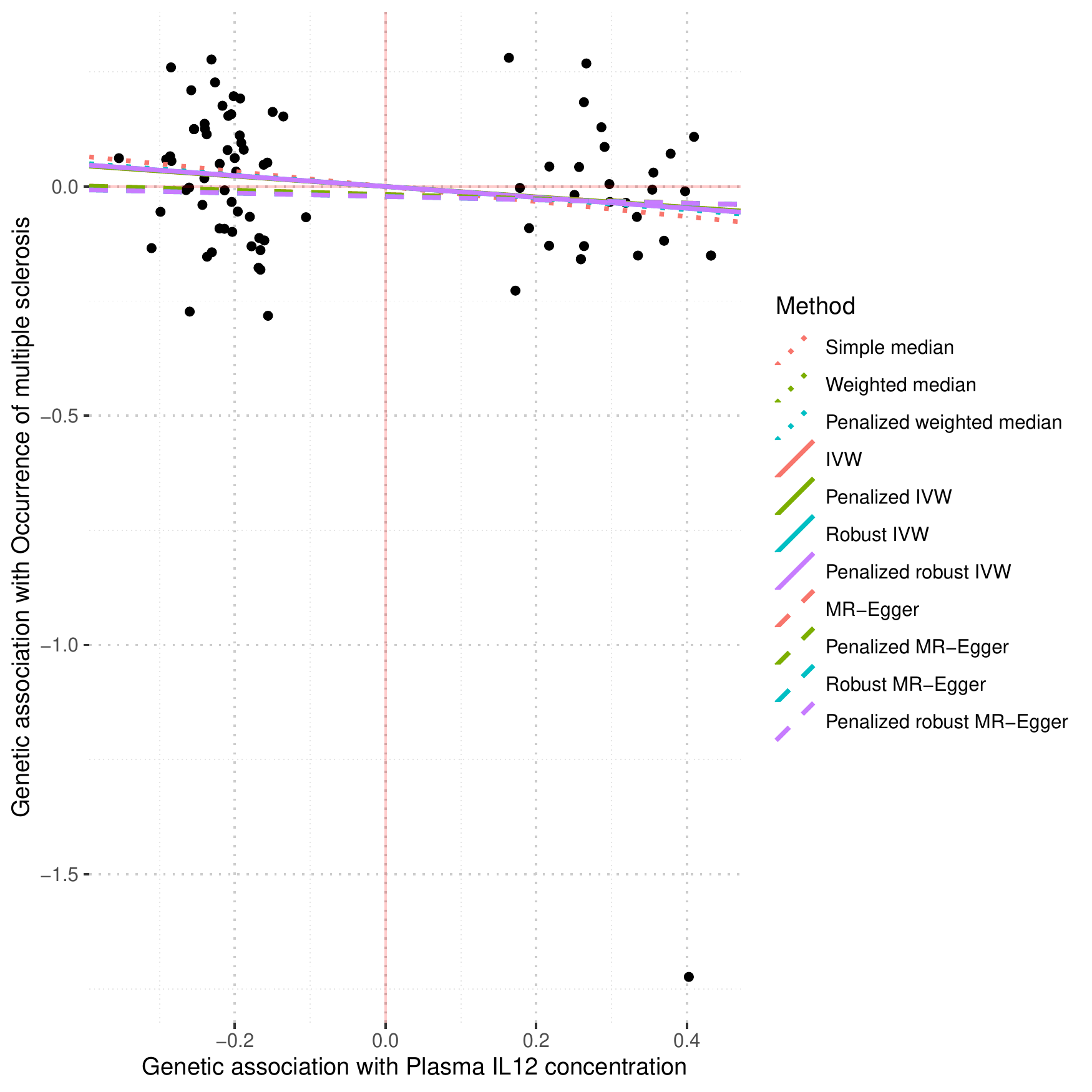} \hspace{1.5cm}
\includegraphics[width=0.5\textwidth,height=5cm]{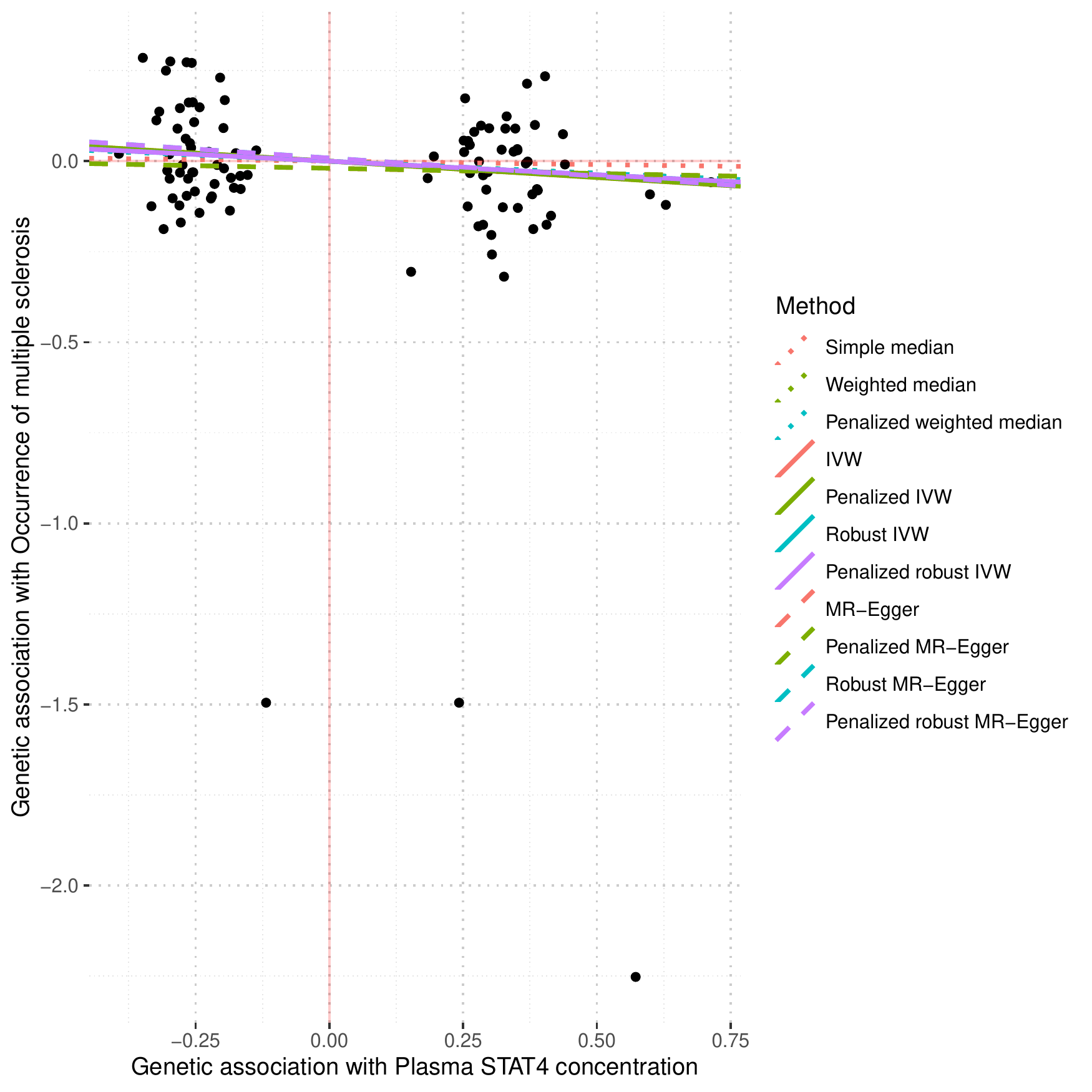}
\caption{\footnotesize Egger plots proteins IL12A and STAT4}
\label{EggerPlotIL12STAT4}
\end{subfigure}

\vspace{0.2cm}

\begin{subfigure}{0.9\textwidth}
\includegraphics[width=0.5\textwidth,height=5cm]{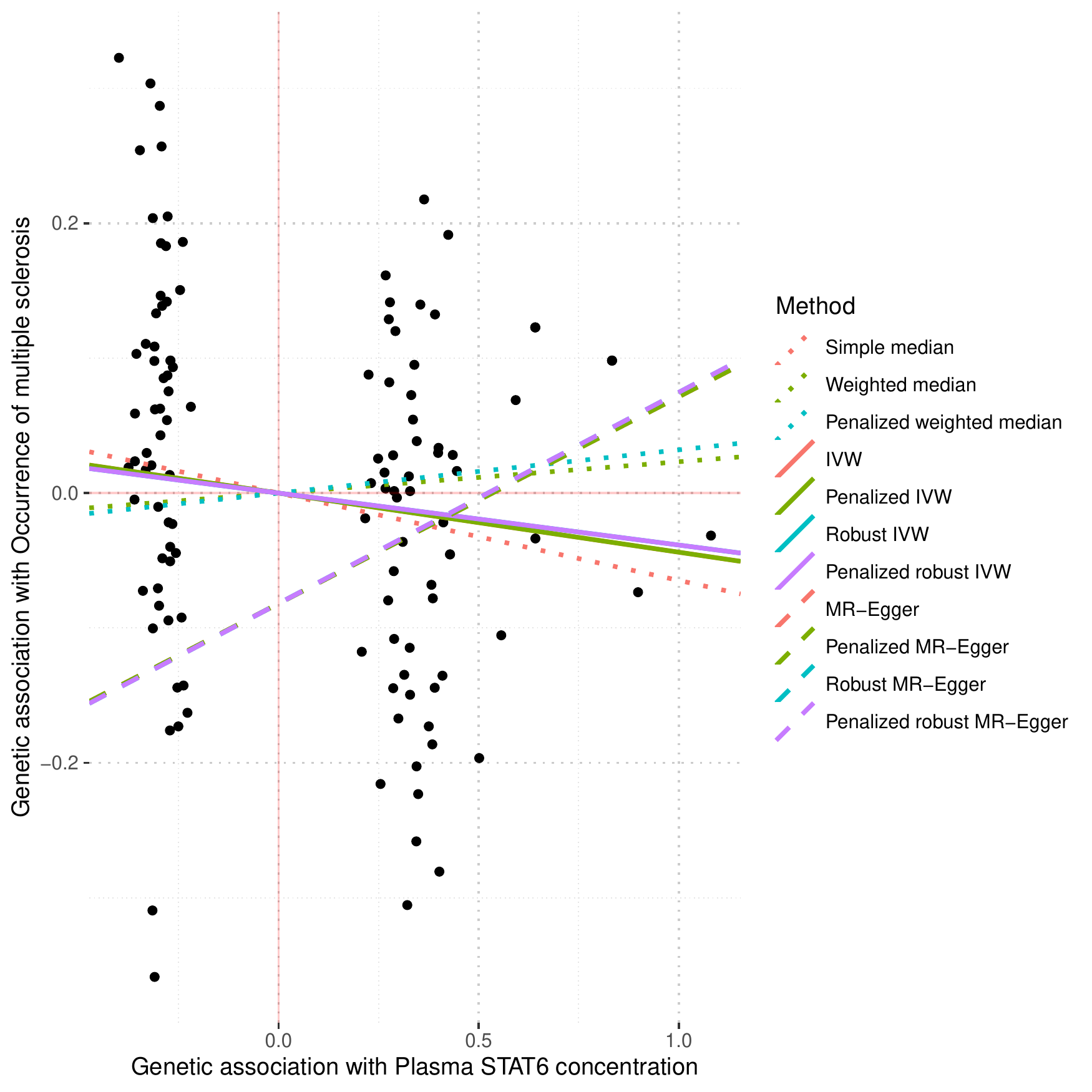}
\label{EggerPlotSTAT6}
\caption{\footnotesize Egger plots protein STAT6}
\end{subfigure}
\end{figure}

\subsection*{Bayesian Analysis}

\vspace{0.4cm}

\noindent In our Bayesian analyses, we acknowledged the fact that the protein level values in MS cases may be reverse-affected due to treatment by treating them as missing, and effectively allowing them to be estimated by the MCMC algorithm. There were missing protein values also in some disease-free subjects. The Bayesian method of dealing with missing $X$ values appears to be superior to simple replacement of those values with the corresponding genotype-based predictions, in two respects. Firstly, within a Bayesian approach to MR, the imputed $X$ values reflect the information contained in individual's genotypes {\em but} also, in addition, they incorporate relevant information contained in the values of $Y$, via coherent flow of the information all across the model. Secondly, within a Bayesian approach to the problem, the uncertainty involved in the imputation of the missing $X$, and in particular the uncertainty induced by the estimation of the $\alpha$ coefficients, is dealt with in a coherent way. This is a crucial festure in the presence of weak instruments. In fact, during MCMC inference, the missing $X$ values are re-imputed at each new iteration of the algorithm, so as to reflect the uncertainty in all the parameters of the model. In each MCMC analysis, we ran the Markov chain for 9000 iterations, and then used the sampled values generated during the last 4500 iterations. Sampled values were used to approximate the posterior distribution of the quantities of interest.

\newpar Reported in Table \ref{Bayesian} are results from a Bayesian analysis of the causal effect of each protein of interest on susceptibility to MS. The estimated (protein-specific) causal effects are expressed as disease log-odds ratios and disease odds ratios for a standard deviation change in level of circulating protein.
 
\vspace{1.5cm}

\begin{longtable}{l*7{p{1cm}}}
\caption{\small Results from a Bayesian analysis of the causal effects of 
five proteins on susceptibility to MS. The last seven columns
of the table report percentile of the posterior distribution
of  the causal effect.\label{Bayesian}}\\
\rowcolor{maroon!10}
\topmidheader{1}{PROTEIN IL4}
\\
Causal Exposure Log Odds Ratio & -1.37 & -0.77 & -0.47 & -0.15 & 0.25 & 0.74 & 1.61\\ 
Causal Exposure Odds Ratio & 0.25 & 0.46 & 0.62 & 0.86 & 1.29 & 2.09 & 5.01 \\ 

\\
\rowcolor{maroon!10}
\topmidheader{1}{PROTEIN IL23A}
\rowcolor{white}
&Min &0.05&0.25 & 0.5 & 0.75 & 0.95 & max\\ 
\endfirsthead
\caption[]{(continued)}\\
\rowcolor{white}
%\vspace{0.2cm}
&Min &0.05&0.25 & 0.5 & 0.75 & 0.95 & max\\
\endhead
Causal Exposure Log Odds Ratio & -1.24 & -0.61 & -0.29 & -0.04 & 0.19 & 0.51 & 1.17 \\ 
Causal Exposure Odds Ratio & 0.29 & 0.54 & 0.75 & 0.96 & 1.20 & 1.66 & 3.22 \\ 

\\
\rowcolor{maroon!10}
\topmidheader{1}{PROTEIN IL12A}
 Causal Exposure Log Odds Ratio & -5.74 & -3.52 & -2.65 & -2.12 & -1.46 & -0.77 & 0.33 \\ 
Causal Exposure Odds Ratio & 0.00 & 0.03 & 0.07 & 0.12 & 0.23 & 0.46 & 1.39 \\ 

\\
\rowcolor{maroon!10}
\topmidheader{1}{PROTEIN IL12A}
\\
Causal Exposure Log Odds Ratio & -5.55 & -3.52 & -2.48 & -1.94 & -1.41 & -0.60 & 0.27 \\ 
Causal Exposure Odds Ratio & 0.00 & 0.03 & 0.08 & 0.14 & 0.25 & 0.55 & 1.30 \\ 

\\
\rowcolor{maroon!10}
\topmidheader{1}{PROTEIN STAT4}
\\
Causal Exposure Log Odds Ratio & -2.04 & -1.14 & -0.78 & -0.62 & -0.50 & -0.09 & 0.38 \\ 
Causal Exposure Odds Ratio & 0.13 & 0.32 & 0.46 & 0.54 & 0.61 & 0.91 & 1.46 \\ 

\\
\rowcolor{maroon!10}
\topmidheader{1}{PROTEIN STAT6}
\\
Causal Exposure Log Odds Ratio & -1.05 & -0.34 & -0.25 & 0.04 & 0.24 & 0.46 & 1.08 \\ 
  Causal Exposure Odds Ratio & 0.35 & 0.71 & 0.78 & 1.04 & 1.27 & 1.59 & 2.93 \\

\end{longtable}

\subsection*{Interpretation of Results}

\noindent Note that Table \ref{Frequentist} contains two entries for IL4 and two entries for IL12. In both cases, the first entry refers to an analysis with the full set of instruments, whereas the second refers to an analysis where one outlier instrumental SNP (with an extreme negative value for its effect on the outcome) was discarded from analysis. For all the five proteins under study, the estimated values for the causal effect of interest (and in some proteins even their signs) were not consistent across the array of frequentist MR algorithms. A discussion of this issue is not within the scope of this paper. An overall tendency of the frequentist estimates of the causal effect to have a negative sign for proteins IL12 and STAT4 is evident.

\newpar The Bayesian analysis highlights a potential causal effect of the circulating concentrations of proteins IL12 and STAT4 on susceptibility to MS, corresponding to the 95\%
credible intervals of these two estimated effects being entirely located in the negative real semi-axis. We conclude that, under the appropriate assumptions,  there is some evidence that proteins IL12 and STAT4, but not the remaining ones, exert a causal effect on risk of MS. The negative sign of the estimates in both cases suggests that an increase on plasma concentration of the protein tends to reduce the risk of MS.

\section*{Discussion}

\subsection*{Methodological}

\noindent In this paper, the Bayesian MR framework of Berzuini and colleagues \cite{Berzuini2018a} has been extended in response to difficulties encountered in the study of an isolated and genetically homogeneous subpopulation of Sardinia with extremely high incidence of MS. Among the difficulties we mention the (inevitably) limited sample size, the strong family relationships between sampke individuals, and the presence of missing exposure values. The missing values arose from the need to acknowledge that exposure values in the diseased may have been reverse-affected by treatment, and from our consequent decision to treat them as missing. Further difficulties arose from the peculiar choice of exposure variable -- the concentration of a protein -- which operates at the beginning of the biological cascade, thereby making it hard to gather strong instrumental information. Instrumental weakness does, in turn, introduce vulnerability to confounding, which is only partly remedied by use of pedigree data (see later comments). 

\newpar The above described, problematic, situation is naturally and conveniently tackled via a Bayesian approach to MR. And for a number of reasons. The first reason being a probabilistically coherent handling of missing data. The second being coherent handling of parameter estimation uncertainty, especially in relation to estimation of weak instrument-exposure associations. The third reason being full exploitation of data information. The fourth being ease of model extension, illustrated in our study by incorporation of between-individual outcome correlation and family effects.The fifth being freedom from large sample asymptotic approximations, which is an advantage in the study of small populations, considering also that introduction  of kinship information in the model will reduce the number of "effective" individuals. Failure to account for this phenomenon, and for family effect, may lead to overoptimistic conclusions about the causal effect of interest.

\newpar This paper illustrates the extension of the Bayesian MR framework to deal with family data. Family data analysis is more robust to population stratification and heterogeneity than analysis of unrelateds, and promises to disentangle inheritable from environmental effects. A potentially fruitful idea is to collect data from unrelated individuals and then to collect further data from the  parents of those individuals, for a joint analysis of the two data  sources. Such a joint analysis can be performed via our proposed  approach by treating parent-child triads as "families". Or one  could use information from previous analyses of unrelateds in  order to shape informative priors for an analysis of pedigree data along our proposed lines. Pedigree analysis might prove  an invaluable tool for studying disease mechanism peculiarities  of small, possibly native and isolated, populations. We are,  in particular, thinking of small populations characterized by  maverick disease patterns, that suffer from inadequate attention  from the medical research community, perhaps outside the western "white" world.

\subsection*{Substantive}

\newpar MS is an immune-mediated demyelinating disease, showing CNS lymphocyte infiltration, production of pro-inflammatory cytokines, and inappropriate activation of Th1, Th17, B, 
and natural killer (NK) cells \cite{vonEssen2019}. Genome-wide association studies discovered, among others, MS-associated risk alleles in the IL12/STAT4 and IL23/STAT3 
pathways implicated in the differentiation of Th1, Th2, and Th17 cells \cite{IMSGC2019}. These pathways are profoundly interlaced, 
considering both the repertoire of protein subunits taking part in the formation of cytokines and of their receptors, as well as the involvement of downstream transcription 
factors that result activated (i.e., signal transducer and activator of transcription, STAT, proteins). In particular, the Interleukin 12 (IL12) family of cytokines is composed 
of four different members (IL12, IL23, IL27, IL35), and each member is a heterodimer composed of two subunits, an $\alpha$-subunit (p35, p19, or p28, encoded by the IL12A, 
IL23A, and IL27A gene, respectively) and a $\beta$-subunit (p40 or Ebi3, encoded by the IL12B and EBI3 gene, respectively) \cite{Vignali2012}. The three $\alpha$ 
subunits are structurally related, and each can pair with either of the structurally homologous $\beta$ subunits. So that, pairing of the $\alpha$-subunits, p35 or p19, with p40, 
generates the two pro-inflammatory IL12 and IL23 cytokines, respectively; conversely, the two immune-suppressive members of the family, i.e. IL27 and IL35, derives from the pairing 
of p28 or p35 with Ebi31 \cite{Vignali2012}. Figure 3 summarizes the composition of the IL12 family, and recapitulates the major links present among cytokines, 
immunologically relevant cells, and MS.

\newpar In our work, we found altered levels of the IL12/STAT4 axis. IL12 is produced mainly by antigen presenting cells (APCs, e.g. dendritic cells or macrophages). It acts as an immunological playmaker by inducing: i) Th1 cell differentiation from CD4+ naive T cells; ii) interferon gamma (IFN-$\gamma$) production; and iii) tumor necrosis factor-alpha (TNF-$\alpha$) production from T and NK cells \cite{Aslani2017}. The hypothesised causal effect of IL12 on MS risk might be mediated by higher levels of expression of this cytokine, with the subsequent IL12-induced production of IFN-$\alpha$. Indeed, IFN-$\gamma$ is a major cytokine found in MS lesions, and it has been found that its levels are greatly increased during MS activity \cite{Lees2007}. A very recent meta-analysis, performed on a total of 226 studies with 13,526 MS patients, seems to confirm that increased levels of IL12 are present in blood and/or cerebrospinal fluid of MS patients \cite{Bai2019}. On the contrary, we found that an increase in plasma concentration of the IL12 protein tends to reduce the risk of MS. In our opinion, this is not surprisingly for a number of reasons.

\newpar First, in the above mentioned meta-analysis, increased levels of IL12 were reported as measured either considering the p40 (IL12B) subunit alone, or the p70 (IL12A+IL12B) heterodimeric protein, whereas our study points to decreased levels in MS of the p35 subunit (IL12A) only. This is not trivial, considering that the p35 subunit is part not only of the IL12 complex, but also of the IL35 cytokine, and, similarly, p40 takes part to the composition of both IL12 and IL23. But while IL12 and IL23 are pro-inflammatory cytokines, IL35 is inhibitory, so that their overall balance is crucial for the modulation of immune function.

\newpar Second, it has been recently demonstrated that IL12-p35 induces expansion of IL10 and IL35-expressing regulatory B cells, thus ameliorating autoimmune disease \cite{Wang2014} \cite{Dambuza2017}.

\newpar Third, also genetics can help us in differentiating the roles for p35 and p40 (rather than IL12 ``as a whole”). In this frame, an interesting review by \cite{vanWanrooij2012} clearly showed that, on the basis on association studies between autoimmune diseases with the various IL12 regions, two major clusters of diseases can be distinguished: the first one including Crohn's disease, ulcerative colitis, psoriasis and psoriatic arthritis, ankylosing spondylitis, and rheumatoid arthritis; these diseases show preferential associations with the IL12B gene region, indicating a pivotal role for the Th17/Th1 pathways. Instead, the second cluster encompasses primary biliary cirrhosis, celiac disease, Graves disease, and MS; these conditions show significant associations with polymorphisms in the IL12A gene region, thus indicating a specific role for p35.
Finally, our data, together with those mentioned above and reported by \cite{Dambuza2017}, might reconcile with the notion that IL12/IL23 antibody therapy failed to be effective in MS \cite{Longbrake2009}.

\vspace{2cm}

\begin{center}
\begin{figure}[!h]
\label{Pathway}
\vspace{0.4cm}
\begin{center}
\scalebox{1.8}{
\includegraphics[width=0.45\textwidth]{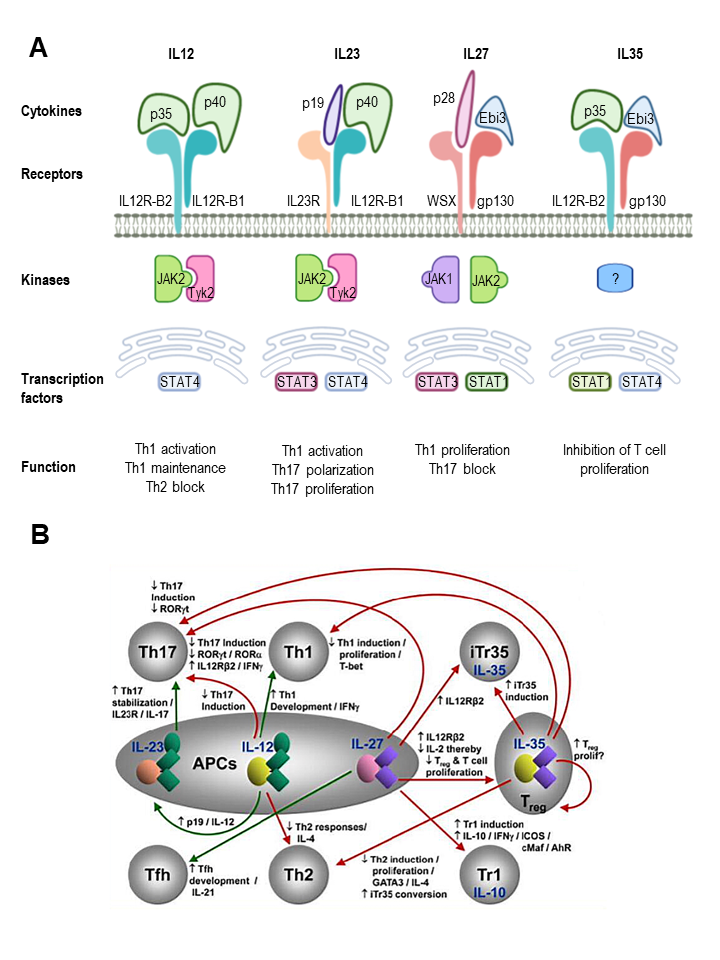}
}
\end{center}
\caption{\footnotesize IL12 family cytokines as a
putative immunological link between IL12A and MS.}
\end{figure}
\end{center}

\section*{Acknowledgements}
We thank the patients, their relatives, and all the volunteers who donated their samples. 
We gratefully acknowledge Dr. Valeria Saddi, Dr. Marialuisa Piras, and all the ascertainment team in Nuoro. 

\section*{Funding}
This work was supported by Fondazione Italiana Sclerosi Multipla [grant number 2009//R//2] and Fondazione Cariplo [grant number 2009-2528].

\section*{Conflict of interest}
None declared

\vspace{1cm}

\bibliographystyle{plain} 
\bibliography{BIBLIOGRAFIA}

\end{document}